\documentclass[final,5p,times,twocolumn]{elsarticle}

\usepackage{latexsym}
\usepackage{amssymb, amsmath, bm, physics}
\usepackage{subcaption}
\usepackage{mathptmx}
\usepackage{flushend}
\usepackage[colorlinks,citecolor=blue,urlcolor=blue,linkcolor=blue]{hyperref}
\usepackage{aas_macros}
\usepackage{orcidlink}

\usepackage{xcolor}

\biboptions{sort&compress}

\begin{document}

\begin{frontmatter}




\title{Quantum corrected thermodynamics and horizon quantization of the Reissner--Nordström black hole}


\author[1]{S. Jalalzadeh \corref{cor1} \orcidlink{0000-0003-4854-2960}}
\ead{shahramjalalzadeh@iyte.edu.tr}
\address[1]{Izmir Institute of Technology, Department of Physics, Urla, 35430, Izmir, Turkiye}
\cortext[cor1]{Corresponding author}
\author[2]{H. Moradpour
\orcidlink{0000-0003-0941-8422}}
\ead{h.moradpour@riaam.ac.ir}
\address[2]{Research Institute for Astronomy and Astrophysics of Maragha (RIAAM), University of Maragheh, Maragheh, 55136-553, Iran}

\begin{abstract}
In this letter, we develop a unified semiclassical framework for the thermodynamics and quantization of the Reissner--Nordstr\"om (RN) black hole (BH) based on the Misner--Sharp--Hernandez (MSH) mass. Treating the quasi-local horizon energies as the relevant thermodynamic variables, we formulate a horizon-by-horizon first law and Smarr relation. Using a reduced phase-space quantization, we obtain a discrete MSH mass spectrum for both horizons, which reproduces the minimal entropy spacing. Quantum transitions between adjacent levels yield Planck-scale corrections to the Hawking temperatures and a universal logarithmic contribution to the entropy, consistent with independent approaches to quantum gravity. We encode these corrections into a quantum-deformed RN geometry via a simple multiplicative factor that preserves the classical horizon positions while reproducing the corrected surface gravities. The associated effective stress tensor behaves as a conserved vacuum-polarization source with characteristic $r^{-4}$ falloff and a small trace, providing a compact representation of semiclassical backreaction. The deformation slightly lowers both horizon temperatures, weakens the inner-horizon instability, and induces tiny shifts in photon-sphere and shadow observables for macroscopic BHs. 
\end{abstract}


\end{frontmatter}

\section{Introduction}
BH thermodynamics reveals a deep interplay between gravitation, quantum theory, and statistical mechanics. Hawking radiation endows BHs with temperature and entropy \cite{Hawking:1975vcx}, motivating the search for a microscopic origin of the Bekenstein--Hawking area law. Among several approaches, canonical quantization in a reduced phase space provides a transparent route by mapping horizon dynamics to a quantum-mechanical system with discrete spectra.

For spherically symmetric spacetimes, the MSH mass \cite{Misner:1964je, Hernandez:1966zia} offers a natural quasi-local energy that underlies the ``unified first law'' of horizon thermodynamics \cite{Hayward:1998ee, Padmanabhan:2002sha}. Evaluated at a Killing horizon, the MSH mass coincides with the thermodynamic energy entering the first law. While this structure is well known for the Schwarzschild case, charged BHs introduce a second (inner) horizon with opposite-sign surface gravity, calling for a framework that treats both horizons on equal footing.

In this letter, we construct such a framework for the RN geometry. First, we show that the MSH energy on each horizon, $M_i=r_i/(2G)$, yields a clean horizon-by-horizon first law with the electromagnetic sector contributing a local work term. The corresponding Smarr relation follows from Euler homogeneity. Next, we implement a reduced phase-space quantization using canonical variables $(M_i,\Pi_i)$ and a transformation to oscillator pairs. The resulting Hamiltonians produce a discrete MSH mass spectrum for both horizons, naturally reproducing the minimal entropy spacing and providing a microscopic basis for horizon thermodynamics \cite{Das:2002xb, Barvinsky:2001tw, Jalalzadeh:2021gtq, Jalalzadeh:2026ssq}.

{Unlike approaches based on the ADM or Komar mass, our framework resolves the thermodynamics
\emph{horizon by horizon}: each Killing horizon obeys a local first law and Smarr relation written
in terms of the corresponding quasi-local MSH energy. At the same time, the two horizons are
\emph{not} independent subsystems in the sense of unconstrained variations, since they remain
coupled through the global parameters $(m,q)$ of the RN family.}
This dual-horizon formulation naturally accommodates the negative surface gravity of the inner (Cauchy)
horizon and allows a unified extension of the reduced phase-space quantization method previously developed
for the Schwarzschild case \cite{Das:2002xb,Barvinsky:2001tw}.
In this way, the quasi-local MSH energy bridges classical horizon thermodynamics, canonical quantization,
and semiclassical backreaction within a single self-consistent setting.

We then analyze quantum transitions between adjacent levels and derive leading corrections to the classical thermodynamics: the temperatures of both horizons are reduced by a small, radius-dependent factor, and the entropy acquires a universal logarithmic term. These findings agree with independent derivations of logarithmic entropy corrections from conformal field theory, loop effects, and general statistical arguments \cite{Kaul:2000kf, Carlip:2000nv, Das:2001ic, Sen:2012dw, Zhu:2008cg}. To geometrize the corrections, we introduce a simple multiplicative deformation of the RN metric function that preserves the horizon locations and asymptotic flatness while reproducing the corrected surface gravities. The corresponding effective stress tensor behaves as a conserved vacuum-polarization source with $r^{-4}$ falloff and a small, nonzero trace, consistent with semiclassical expectations.

Altogether, our analysis links discrete horizon microstructure, corrected thermodynamic relations, and semiclassical backreaction within a single covariant construction based on the MSH mass. The framework reduces smoothly to classical RN in the macroscopic limit and suggests controlled modifications to inner-horizon stability, evaporation, and light-ring observables. We close by outlining extensions to rotating and cosmological BHs and potential observational implications for near-extremal and primordial BHs. In this work, we adopt natural units, in which $k_B=c=\hbar=1$.

\section{Thermodynamics of the Reissner--Nordström Black Hole from the Misner--Sharp--Hernandez Mass }
\label{sec:MSHthermo}

The well-known line element of the RN is given by
\begin{equation}\label{1-1}
\mathrm ds^2=-f(r)\mathrm dt^2+\frac{\mathrm dr^2}{f(r)}+r^2 \mathrm d\Omega_2^2,
\quad
f(r)=1-\frac{2Gm}{r}+\frac{G q^2}{r^2},
\end{equation}
where $\mathrm d\Omega_2^2$ is the line element of the standard unit sphere, $m$ and $q$ are the ADM mass and electric charge, respectively.\footnote{We adopt the standard Einstein--Maxwell normalization used in, e.g., \cite{Hayward:1998ee,Padmanabhan:2002sha}.}
The inner (or Cauchy) ($-$) and outer ($+$) horizons are at the roots of $f(r)=0$, i.e.,
\begin{equation}\label{Horizons}
r_\pm = Gm \pm \sqrt{G^2 m^2 - G q^2}.
\end{equation}
Through the horizon Eq. (\ref{Horizons}), we have the following relations
\begin{equation}
    \label{Horizon2}
    r_+ + r_- = 2Gm,\quad r_+ r_- = G q^2 .
\end{equation}

The first law of thermodynamics for the outer horizon is given by
\begin{equation}\label{First1}
    \delta m=T_+\delta S_++\frac{q}{Gr_+}\delta q,
\end{equation}
where $T_+$ and $S_+$ are the Hawking temperature and the 
Bekenstein--Hawking entropy of the outer horizon, respectively.
The authors of Refs. \cite{Das:2002xb, Barvinsky:2000gf} utilized the above form of the first law of thermodynamics to define canonical conjugate variables for the outer horizon to quantize its surface area. In this article, our first aim is to extend the formalism of the mentioned references into both horizons. To achieve this aim, we reformulate the first law of thermodynamics in terms of the MSH mass of two horizons.

For any spherically symmetric spacetime of the form of (\ref{1-1}), the MSH mass is \cite{Misner:1964je, Hernandez:1966zia, Hayward:1998ee}
\begin{equation}
M_{\rm MSH}(r)=\frac{r}{2G}\!\left(1-g^{ab}\nabla_a r\nabla_b r\right)
=\frac{r}{2G}\big(1-f(r)\big)
= m-\frac{q^2}{2r}.
\end{equation}
Evaluated on a horizon $r=r_i$ $(i=\pm)$ with $f(r_i)=0$, we find
\begin{equation}
M_i \equiv M_{\rm MSH}(r_i)=m-\frac{q^2}{2r_i}=\frac{r_i}{2G}.
\label{eq:MSHmassG}
\end{equation}
Thus, the quasi-local thermodynamic energy at each horizon is $M_i=r_i/(2G)$.

Now, let us remind two horizons thermodynamical variables
\begin{equation}\label{2-77}
\begin{split}
S_i&=\frac{A_i}{4G}=\frac{\pi r_i^2}{G},\qquad
V_i=\frac{4\pi}{3}r_i^3,\\
T_i&=\frac{\kappa_i}{2\pi},\qquad
\kappa_i=\frac{r_i-r_j}{2r_i^2}\ (j\neq i).
\end{split}
\end{equation}
where $S_i$ is the Bekenstein--Hawking entropy, $V_i$ is the thermodynamical volume, $T_i$ is the corresponding temperature, and $\kappa_i$ is the surface gravity of the i-th horizon. 

For the matter (Maxwell) sector, we use the local electromagnetic energy density at the horizon,
\begin{equation}
p_i=\frac{q^2}{8\pi r_i^4}.
\end{equation}
Then, one can rewrite the first law (\ref{First1}) for both horizons in the following compact form
\begin{equation}
 \delta M_i = T_i\,\delta S_i + p_i\,\delta V_i
\ =\ T_i\,\delta S_i+\frac{q^2}{8\pi r_i^{4}}\,\delta V_i.
\label{eq:firstlawMSH_G}
\end{equation}
At fixed $q$, one checks from $f(r_i)=0$ that $2G\,\delta m=(1-Gq^2/r_i^2)\,\delta r_i$, which yields $\delta M_i=\delta(r_i/2G)=\frac{1}{2G}\delta r_i$; the right-hand side of \eqref{eq:firstlawMSH_G} reduces to the same quantity using $r_ir_j=Gq^2$.

Under $r\!\to\!\lambda r$, one has $S_i\!\sim\!\lambda^2$, $V_i\!\sim\!\lambda^3$, $p_i\!\sim\!\lambda^{-4}$, and $M_i\!\sim\!\lambda$.
Euler homogeneity gives
\begin{equation}
M_i = 2 T_i S_i + 3 p_i V_i.
\label{eq:Smarr_horizon}
\end{equation}
For RN, $3p_iV_i=q^2/(2r_i)=\Phi_i q/2$ with $\Phi_i=q/r_i$, so equivalently
\begin{equation}
M_i = 2 T_i S_i + \frac{1}{2}\,\Phi_i q .
\end{equation}
Using $r_ir_j=Gq^2$, Eq. \eqref{eq:Smarr_horizon} reproduces $M_i=r_i/(2G)$ identically. The Smarr relation \eqref{eq:Smarr_horizon} follows directly from Euler homogeneity of degree one in the radial coordinate $r_i$, a result consistent with the horizon thermodynamics of Padmanabhan \cite{Padmanabhan:2002sha}.  
Each horizon satisfies the Smarr relation, reinforcing the interpretation of the MSH energy as the quasi-local enthalpy of the corresponding region.

Because $\kappa_-<0$, the Cauchy horizon has a negative Hawking temperature,
\begin{equation}
T_-=\frac{\kappa_-}{2\pi}<0 .
\end{equation}
This sign reflects the reversed orientation of the inner horizon’s null generator and is standard in the ``first law of inner mechanics'' \cite{Castro:2012av, Chen:2012mh}.  
One may alternatively use $|T_-|=-T_-$ and write the law as
\begin{equation}
\delta M_- = -\,|T_-|\,\delta S_- + p_-\,\delta V_- ,
\end{equation}
which is completely equivalent to \eqref{eq:firstlawMSH_G}.

With $M_i=r_i/(2G)$ and
\begin{equation}\label{Temp}
T_i=\frac{r_i^2-Gq^2}{4\pi r_i^3},
\end{equation}
the heat capacity at constant $q$ is
\begin{equation}
 C_i:=\left(\frac{\partial M_i}{\partial T_i}\right)_q 
= \frac{2\pi r_i^{4}}{\,G\,(3Gq^2-r_i^2)\,} .
\end{equation}
For $r_i^2=3Gq^2$, the heat capacity diverges, signaling a Davies-type phase transition.  
The outer horizon is thermodynamically stable ($C_+>0$) in the interval $q\sqrt{G}<r_+<\sqrt{3G}q$ and unstable ($C_+<0$) for $r_+>\sqrt{3G}q$, while the inner horizon exhibits the opposite behavior.  
These features mirror the characteristic thermodynamic instability of charged black holes in asymptotically flat spacetimes.

Eqs. \eqref{eq:firstlawMSH_G} and \eqref{eq:Smarr_horizon} show that the MSH energy \emph{on the horizon} plays the role of the thermodynamic energy, while the electromagnetic sector contributes a local work term $p_i\,\delta V_i$.
This realizes Hayward’s ``unified first law'' and Padmanabhan’s horizon thermodynamics in a two-horizon setting \cite{Hayward:1998ee, Padmanabhan:2002sha, Akbar:2006kj}, and it dovetails with the inner-horizon mechanics literature \cite{Castro:2012av, Chen:2012mh}.

\section{Quantization and MSH mass spectrum}\label{3}

Since the mass $m$ and charge $q$ are the only time-independent, coordinate-invariant observables in spherically symmetric geometrodynamics, a reduced action can be written as
\begin{equation}\label{2-1}
    \mathcal I=\int\!\left\{\dot m\,P_m+\dot q\,P_q-H(m,q)\right\}\mathrm dt,
\end{equation}
where $P_m$ and $P_q$ are the conjugate momenta. This generalizes the Schwarzschild reduced action to charged BHs \cite{Barvinsky:2001tw, Barvinsky:2000gf}. Because $m$ and $q$ are constants of motion, $H$ is independent of their conjugate momenta.

It is convenient to trade $(m,q)$ for the MSH horizon energies $(M_+,M_-)$, with conjugate variables $(\Pi_+,\Pi_-)$. The action \eqref{2-1} becomes
\begin{equation}
    \label{2-2}
    \mathcal I=\int\!\left\{\sum_{i=-}^{+}\dot M_i\,\Pi_i-H(M_i)\right\}\mathrm dt.
\end{equation}
\noindent
{We stress that $(M_+,M_-)$ are not introduced as additional degrees of freedom; rather, they constitute
an invertible reparameterization of the same $2D$ reduced phase space spanned by $(m,q)$.
Using $r_+ + r_- = 2Gm$ and $r_+ r_- = Gq^2$ together with the horizon identity $M_i=r_i/(2G)$,
we obtain
\begin{equation}
m = M_+ + M_-,
\qquad
q^2 = 4G\,M_+ M_-,
\label{eq:m_q_in_terms_of_Mpm}
\end{equation}
(up to the sign of $q$). Hence $(m,q)$ and $(M_+,M_-)$ are related by an algebraically invertible mapping,
and $(M_+,M_-)$ provides a valid coordinate chart on the same reduced phase space.}

{Finally, the conjugate momenta $(\Pi_+,\Pi_-)$ are defined by demanding that the canonical one-form
(and hence the symplectic structure) be preserved under the change of variables. Concretely, we impose
\begin{equation}
P_m\,dm + P_q\,dq \;=\; \Pi_+\,dM_+ + \Pi_-\,dM_-,
\label{eq:symplectic_match_reply}
\end{equation}
so that the transformation $(m,q;P_m,P_q)\mapsto(M_+,M_-;\Pi_+,\Pi_-)$ is canonical and therefore preserves
Poisson brackets. Using the invertible relations
\begin{equation}
m = M_+ + M_-,
\qquad
q^2 = 4G\,M_+ M_-,
\label{eq:mq_in_terms_of_Mpm_reply}
\end{equation}
one has
\begin{equation}
dm = dM_+ + dM_-,
\qquad
dq = \frac{2G}{q}\big(M_-\,dM_+ + M_+\,dM_-\big),
\label{eq:dm_dq_in_terms_of_dM_reply}
\end{equation}
(for $q\neq 0$; the neutral limit can be taken smoothly after the final expressions are obtained).
Substituting \eqref{eq:dm_dq_in_terms_of_dM_reply} into the left-hand side of
\eqref{eq:symplectic_match_reply} and collecting the coefficients of $dM_\pm$, we obtain
\begin{equation}
\Pi_+ \;=\; P_m + \frac{2G M_-}{q}\,P_q,
\qquad
\Pi_- \;=\; P_m + \frac{2G M_+}{q}\,P_q.
\label{eq:Pi_pm_explicit_reply}
\end{equation}
Equivalently, one may write these relations in terms of the horizon radii using $M_\pm=r_\pm/(2G)$.
By construction, Eq.~\eqref{eq:symplectic_match_reply} implies
\begin{equation}
\{M_i,\Pi_j\}=\delta_{ij},
\qquad
\{M_i,M_j\}=0,
\qquad
\{\Pi_i,\Pi_j\}=0,
\label{eq:PB_canonical_reply}
\end{equation}
and thus the use of $(M_+,M_-;\Pi_+,\Pi_-)$ in the reduced action is a canonical reparameterization
of the same $2D$ reduced phase space rather than an introduction of additional degrees of freedom.}

Since $H$ is momentum independent, the $\Pi_i$ are proportional to the (Euclidean) time coordinate. Regularity of the Euclidean section (absence of conical defects), therefore, imposes the standard periodicity \cite{Das:2002xb, Barvinsky:2000gf}
\begin{equation}
    \label{2-3}
    \Pi_i\sim \Pi_i+\frac{1}{T_i},
\end{equation}
where $T_i$ is the (signed) Hawking temperature of the corresponding Killing horizon.

Following \cite{Das:2002xb}, impose this periodicity via a canonical transformation
\begin{equation}
    \label{2-4}
    x_i=\sqrt{B_i}\,\cos(\kappa_i\Pi_i),\qquad
    \pi_i=\sqrt{B_i}\,\sin(\kappa_i\Pi_i),
\end{equation}
with $\kappa_i$ the surface gravity. Using $\{M_i,\Pi_i\}=1$ and the horizon first law in the MSH form,
\(
\kappa_i=8\pi G\,\partial M_i/\partial A_i
\)
(see Sec.~\ref{sec:MSHthermo}), one finds
\begin{equation}
\{x_i,\pi_i\}=\frac{\kappa_i}{2}\frac{\partial B_i}{\partial M_i}
=4\pi G\,\frac{\partial B_i}{\partial A_i}=1
\quad\Longrightarrow\quad
B_i=\frac{A_i}{4\pi G}=4M_i^2G.
\end{equation}
Therefore, we have
\begin{equation}
\sqrt{B_i}=2\sqrt{G}M_i.
\end{equation}
and 
\begin{equation}
    \label{2-5}
    M_i^2=\frac{m_{\rm P}^2}{4}\,\big(x_i^2+\pi_i^2\big),
\end{equation}
where $m_{\rm P}=1/\sqrt{G}$ is the Planck mass (natural units).

Canonical quantization in the $x_i$–representation,
\(
\pi_i\mapsto -i\,\mathrm d/\mathrm dx_i,
\)
turns \eqref{2-5} into two decoupled harmonic-oscillator equations,
\begin{equation}
    \label{2-6}
-\frac{1}{2}\frac{\mathrm d^2\psi_i}{\mathrm dx_i^2}
+\frac{1}{2}x_i^2\,\psi_i
= \frac{2M_i^2}{m_{\rm P}^2}\,\psi_i.
\qquad i=-,+,
\end{equation}
The MSH mass spectrum then follows:
\begin{equation}\label{2-7}
   M_i=\frac{m_{\rm P}}{\sqrt{2}}\,
   \sqrt{n_i+\frac{1}{2}},
   \quad n_i=0,1,2,\ldots\, .
\end{equation}
This reproduces the Schwarzschild result as a one-horizon limit and generalizes earlier area/mass quantization analyses to the two-horizon RN case \cite{Das:2002xb, Jalalzadeh:2022rxx, Louko:1996md,2004PhLB226X}. The normalized eigenfunctions are the Hermite polynomials,
\[
\psi_{n_i}(x_i)=\frac{H_{n_i}(x_i)}{\pi^{1/4}\sqrt{2^{n_i}n_i!}}\,e^{-x_i^2/2},
\]
which form a complete orthonormal basis for the horizon’s quantum configuration space.

\section{Entropy and logarithmic correction term}\label{4}
In the RN spacetime, the static Killing vector $\xi^a=\partial_t^a$ changes its causal character across the horizons: it is timelike in the region between the outer and inner horizons, null on each horizon, and spacelike inside the Cauchy region.  
This flip of signature implies that the associated Killing energy $E=-p_a\xi^a$ reverses sign when an observer crosses a horizon.  
During quantum pair creation near the horizons, the negative-energy partner falls inward while its positive-energy counterpart escapes outward, an analogue of the Penrose process in charged black holes.  
As a result, the quasi-local MSH energy decreases at the outer horizon but increases at the inner horizon.  
In the discrete spectrum picture \eqref{2-7}, this corresponds to transitions $n_i\!\to\! n_i\mp1$ (upper sign for the inner horizon, lower sign for the outer one), so that
\begin{equation}
\label{4-1}
\Delta M_i=\sigma_i\!\left[M_i(n_i)-M_i(n_i+1)\right],\quad
\sigma_\pm=\mp1,
\end{equation}
which in the macroscopic limit ($n_i\gg1$) becomes
\begin{equation}\label{4-1aaa}
\Delta M_i\simeq\sigma_i\,\frac{m_{\rm P}^2}{4M_i}
\!\left(1-\frac{m_{\rm P}^2}{8M_i^2}\right).
\end{equation}
This convention ensures that $\Delta M_+<0$ (mass loss through Hawking emission) and $\Delta M_->0$ (energy inflow via negative Killing energy), fully consistent with the reversed orientation of the inner horizon generator.

{Inserting the minimal jumps into the horizon first law (MSH form) (\ref{eq:firstlawMSH_G}), we find a difference form of the first law:
\begin{equation}\label{eq:DiscFirstLaw}
\Delta M_i= T_i\,\Delta S_i + p_i\,\Delta V_i,
\qquad
p_i=\frac{q^2}{8\pi r_i^4}.
\end{equation}
Now, utilizing the relations 
\begin{equation}
r_i=2GM_i,~~~~~V_i=\frac{4\pi r_i^3}{3}=\frac{32\pi G^3M_i^3}{3},
\label{eq:rV_in_terms_of_M}
\end{equation}
one can rewrite the above difference equation as
\begin{equation}
\begin{split}
    \Delta M_i&=T_i\Delta S_i+\frac{q^2}{8 r_i^4}(32G^3M_i^2\Delta M_i)\\
    &=T_i\Delta S_i+\frac{Gq^2}{ r_i^2}\Delta M_i.
    \end{split}
\end{equation}
Thus, 
\begin{equation}
    T_i=\left(1-\frac{Gq^2}{r_i^2}\right)\frac{\Delta M_i}{\Delta S_i}.
\end{equation}}

{To use this transition formula, we need the entropy spacing $\Delta S_i$.
From Sec.~\ref{3} we have $B_i=A_i/(4\pi G)=4GM_i^2$, hence
\begin{equation}
A_i = 16\pi G^2 M_i^2,
\qquad
S_i=\frac{A_i}{4G}=4\pi G\,M_i^2.
\label{eq:S_in_terms_of_Ma}
\end{equation}
Using the spectrum \eqref{2-7}, $M_i^2=\frac{1}{2G}\left(n_i+\frac12\right)$, we have
\begin{equation}
S_i = 2\pi\left(n_i+\frac12\right)
\quad\Longrightarrow\quad
\Delta S_i = 2\pi.
\label{eq:DeltaS_2pia}
\end{equation}
Therefore, one obtains the quantum-corrected temperatures
\begin{equation}
\label{4-3}
T_i^{\rm (Q)}
=\frac{\big(1-\tfrac{Gq^2}{r_i^2}\big)\,\Delta M_i}{\Delta S_i}
\simeq \frac{r_i^2-Gq^2}{4\pi r_i^3}
\left(1-\frac{G}{2r_i^2}\right),
\end{equation}
where we used $Gq^2/r_i^2=r_j/r_i$ and \eqref{4-1aaa}. The leading factor is precisely the classical (signed) Hawking temperature \eqref{Temp}; the bracket multiplies it by a small, Planck-scale correction of order $G/r_i^2$. For the inner (Cauchy) horizon $\kappa_-<0$, hence $T_-<0$; one may equivalently write the law with $|T_-|$ and an overall minus sign, as discussed in Sec.~\ref{sec:MSHthermo}.}

To infer the entropy, use the differential form of the first law with \eqref{4-3} and expand to leading quantum order. Integrating along quasi-static variations yields
\begin{equation}
    \label{4-4}
    S_i
    = \frac{A_i}{4G}
      + \frac{\pi}{2}\,\ln\!\left(\frac{A_i}{4G}\right)
      + \text{const.}
\end{equation}
The logarithmic contribution is the dominant quantum correction at large area and is widely encountered across independent approaches to quantum gravity and horizon thermodynamics \cite{Kaul:2000kf, Carlip:2000nv, Das:2001ic, Sen:2012dw, Zhu:2008cg}. Higher-order terms from the large-$n_i$ expansion in \eqref{4-1aaa} generate further inverse-area corrections, which are subleading for macroscopic horizons.

\section{Quantum-corrected Reissner--Nordstr\"om geometry and semiclassical source}
\label{sec:Qmetric}

{To geometrize the quantum-corrected horizon thermodynamics obtained in Sec.~\ref{4}, we consider the standard
static, spherically symmetric ansatz \eqref{1-1} with a deformed lapse function $f_{(Q)}(r)$.
The quantized transition analysis implies the leading correction to the (signed) Hawking temperature,
and hence to the surface gravity,
\begin{equation}
T_{i}^{(Q)}=T_{i}^{(\mathrm{cl})}\!\left(1-\frac{G}{2r_{i}^{2}}\right),
\qquad
\kappa_{i}^{(Q)}=2\pi T_{i}^{(Q)}
=\kappa_{i}^{(\mathrm{cl})}\!\left(1-\frac{G}{2r_{i}^{2}}\right),
\label{eq:kappa_Q}
\end{equation}
where $T_{i}^{(\mathrm{cl})}$ and $\kappa_{i}^{(\mathrm{cl})}$ are given by \eqref{2-77}. Since for the metric
\eqref{1-1} one has $\kappa_i=\tfrac12 f'(r_i)$, the semiclassical input extracted from thermodynamics fixes the
\emph{near-horizon jet} of $f$ (in practice, its first derivative at $r=r_i$), but does not uniquely determine the
off-horizon radial profile of a quantum-corrected geometry. Consequently, any admissible $f_{(Q)}(r)$ must satisfy
the boundary conditions
\begin{equation}
f_{(Q)}(r_i)=0,
\qquad
\frac12 f_{(Q)}'(r_i)=\kappa_i^{(Q)},
\qquad (i=\pm),
\label{eq:BC_metric}
\end{equation}
together with asymptotic flatness $f_{(Q)}(r)\to 1$ as $r\to\infty$.}

{A generic additive deformation $f_{(Q)}(r)=f_{(\mathrm{cl})}(r)+\delta f(r)$ shifts the horizon locations
$r_i\to r_i^{(Q)}=r_i+\delta r_i$ unless $\delta f(r_i)=0$. Indeed, expanding $f_{(Q)}(r_i^{(Q)})=0$ around $r_i$
yields
\begin{equation}
\delta r_i=-\frac{\delta f(r_i)}{f_{(\mathrm{cl})}'(r_i)}+\cdots,
\label{eq:horizon_shift_generic}
\end{equation}
so the roots generically move at the same perturbative order.
In the present construction, however, the corrected temperatures \eqref{eq:kappa_Q} are obtained for the \emph{same}
Killing horizons $r=r_\pm$ appearing in the classical RN family (equivalently, for fixed global parameters $(m,q)$).
It is therefore natural to adopt a deformation that preserves the classical horizon radii exactly, while modifying
only the near-horizon slope (and hence $\kappa_i$).}

{This can be implemented in a particularly simple way by factorizing the classical lapse,
\begin{equation}
f_{(Q)}(r)=\Psi(r)\,f_{(\mathrm{cl})}(r),
\label{eq:AQ_def}
\end{equation}
with a smooth dimensionless function $\Psi(r)$.
Since $f_{(\mathrm{cl})}(r_i)=0$, the horizon condition is automatically preserved:
$f_{(Q)}(r_i)=\Psi(r_i)f_{(\mathrm{cl})}(r_i)=0$, so $r=r_i$ remain roots of $f_{(Q)}$ \emph{without shifting the
classical horizon radii}. Moreover,
\begin{equation}
f_{(Q)}'(r_i)=\Psi(r_i)\,f_{(\mathrm{cl})}'(r_i)+\Psi'(r_i)\,f_{(\mathrm{cl})}(r_i)
=\Psi(r_i)\,f_{(\mathrm{cl})}'(r_i),
\label{eq:fprime_factor}
\end{equation}
and therefore the surface gravity matching becomes algebraic:
\begin{equation}
\kappa_i^{(Q)}=\frac12 f_{(Q)}'(r_i)=\Psi(r_i)\,\frac12 f_{(\mathrm{cl})}'(r_i)
=\Psi(r_i)\,\kappa_i^{(\mathrm{cl})}.
\label{eq:kappa_match_algebraic}
\end{equation}
Comparing \eqref{eq:kappa_match_algebraic} with \eqref{eq:kappa_Q} yields the horizon constraint
\begin{equation}
\Psi(r_i)=1-\frac{G}{2r_i^{2}}.
\label{eq:Psi_horizon_match}
\end{equation}}

{The horizon conditions \eqref{eq:Psi_horizon_match} and asymptotic flatness $\Psi(\infty)=1$ do not uniquely fix
$\Psi(r)$ away from the horizons. From an effective-field-theory viewpoint, $\Psi(r)-1$ must be a small,
dimensionless function controlled by the Planck length $\ell_{\mathrm P}=\sqrt{G}$; in a static, spherically
symmetric setting the natural expansion parameter is $(\ell_{\mathrm P}/r)^2=G/r^2$. Thus, to leading order one
expects
\begin{equation}
\Psi(r)=1-\alpha\,\frac{G}{r^2}+\mathcal{O}\!\left(\frac{G^2}{r^4}\right),
\label{eq:Psi_EFT_expand}
\end{equation}
with $\alpha$ fixed by the horizon data. Imposing \eqref{eq:Psi_horizon_match} at leading order gives
$\alpha=\tfrac12$, and a simplest analytic interpolation consistent with $\Psi(\infty)=1$ is
\begin{equation}
\Psi(r)=1-\frac{G}{2r^{2}}.
\label{eq:Psi_profile}
\end{equation}
We stress that \eqref{eq:Psi_profile} is adopted as a \emph{representative} minimal profile; any smooth
$\Psi(r)$ satisfying $\Psi(\infty)=1$ and \eqref{eq:Psi_horizon_match} reproduces the corrected horizon temperatures
to the order considered.
In particular, the results of Secs.~\ref{sec:MSHthermo}--\ref{4} (first law/Smarr relations, spectra, entropy spacing, and the logarithmic correction) do not rely on fixing a unique bulk profile for $\Psi(r)$; only the horizon values $\Psi(r_i)$ enter through $\kappa_i=\tfrac12 f'(r_i)$.}

{With \eqref{eq:Psi_profile}, the quantum-corrected RN lapse becomes
\begin{equation}
f_{(Q)}(r)
=\Big(1-\frac{G}{2r^{2}}\Big)
\Big(1-\frac{2Gm}{r}+\frac{Gq^{2}}{r^{2}}\Big),
\label{eq:AQ_final}
\end{equation}
and the corresponding line element is
\begin{equation}
ds^{2}
=-f_{(Q)}(r)\,dt^{2}
+\frac{dr^{2}}{f_{(Q)}(r)}
+r^{2}d\Omega_{2}^{2}.
\label{eq:metric_final}
\end{equation}}

{In the quantum model, the horizon radii satisfy $r_i^{2}=G(2n_i+1)$, implying a minimal spacing
$\Delta r_i\sim G/r_i$. The representative profile \eqref{eq:Psi_profile} varies precisely on this scale,
$\partial_r\Psi\sim G/r^3$, ensuring that the geometric deformation is consistent with the intrinsic quantum
uncertainty of the radial coordinate.}

{At each horizon, the construction guarantees $\kappa_i^{(Q)}=\kappa_i^{(\mathrm{cl})}\big(1-\frac{G}{2r_i^2}\big)$,
so both (signed) temperatures decrease slightly while $r_\pm$ remain unchanged. The spacetime is asymptotically flat since $\Psi\to 1$ as $r\to\infty$. In addition, the deformation induces tiny shifts in the photon sphere and the shadow
observables, scaling as $G/r^2$,
\begin{equation}
r_{(\mathrm{ph})}^{(Q)}=r_{(\mathrm{ph})}^{(\mathrm{cl})}\!\left(1-\frac{G}{4\,r_{(\mathrm{ph})}^{2}}\right),
\qquad
D_{(\mathrm{sh})}^{(Q)}\simeq D_{(\mathrm{sh})}^{(\mathrm{cl})}\!\left(1-\frac{G}{4\,r_{+}^{2}}\right).
\label{eq:photon_shadow_shift}
\end{equation}
and proportionally small corrections to the eikonal quasinormal frequencies. At the Cauchy horizon,
$\kappa_-^{(Q)}<0$ remains negative, while the correction reduces
$|\kappa_-|$ by the factor $\big(1-\frac{G}{2r_-^2}\big)$, which slightly weakens the mass-inflation instability at
leading order.}

{Finally, Eq.~\eqref{eq:metric_final} can be regarded as a semiclassical RN spacetime supported by an effective
quantum energy--momentum tensor $T^{(Q)}_{\mu\nu}$ that encodes vacuum-polarization backreaction and exhibits the
characteristic $r^{-4}$ falloff at leading order. We derive this tensor in the next section.}

{The multiplicative factor introduces an additional zero of $f_{(Q)}$ at a radius of order $\sqrt{G}$, which may be
interpreted as an extra inner horizon in this effective description. Curvature invariants remain finite there within
the semiclassical approximation, while the only true curvature singularity continues to reside at $r=0$, as in the
classical RN solution. Since curvatures become Planckian near $r\sim\sqrt{G}$, the effective metric
\eqref{eq:metric_final} should be regarded as a leading-order semiclassical model with a conservative range of
validity $r\gg \sqrt{G}$; a complete description of the deep interior requires full quantum-gravity dynamics beyond
the present framework.}

{The profile \eqref{eq:Psi_profile} is adopted as a minimal analytic interpolation consistent with the corrected
horizon data and $\Psi(\infty)=1$; it is not intended as a unique determination of the full quantum geometry.
In particular, the multiplicative deformation may introduce additional roots of $f_{(Q)}$ at radii of order
$r\sim \sqrt{G}$, i.e.\ near the Planck scale, where the semiclassical expansion in $G/r^2$ ceases to be
parametrically controlled. Consequently, any interpretation of the deep interior based on \eqref{eq:metric_final}
should be viewed as qualitative. Throughout the regime $r\gg \sqrt{G}$ relevant for semiclassical thermodynamics
and photon-sphere physics, the deformation remains perturbative, and the metric approaches the classical RN form
smoothly. A complete description of the region $r\lesssim \sqrt{G}$ requires the full quantum-gravity dynamics
beyond the present effective model.}


\section{Effective quantum stress tensor at leading order}
\label{subsec:Tq}

The Einstein equations can be written as
\begin{equation}\label{eq:defTQ}
\begin{split}
&G_{\mu\nu}[f_{(\rm Q)}]
=8\pi G\Big(T^{(\rm EM)}_{\mu\nu}+T^{(\rm Q)}_{\mu\nu}\Big),
\\
&T^{(\rm Q)}_{\mu\nu}=\frac{1}{8\pi G}\Big(G_{\mu\nu}[f_{(\rm Q)}]-G_{\mu\nu}[f_{(\rm cl)}]\Big),
\end{split}
\end{equation}
where $T^{(\rm EM)}_{\mu\nu}$ is the standard Maxwell stress tensor.  
For the metric \eqref{eq:metric_final},
\begin{equation}
G^{t}{}_{t}=G^{r}{}_{r}=\frac{f_{(\rm Q)}'r+f_{(\rm Q)}-1}{r^{2}},
\quad
G^{\theta}{}_{\theta}=G^{\phi}{}_{\phi}=\frac{f_{(\rm Q)}''r+2f_{(\rm Q)}'}{2r}.
\label{eq:Einstein_components}
\end{equation}
{Using $f_{(\rm Q)}=\Psi f_{(\rm cl)}$ with $\Psi(r)=1-\frac{G}{2r^{2}}$, $\Psi'(r)=\frac{G}{r^{3}}$, and $\Psi''(r)=-\frac{3G}{r^{4}}$, one finds
\begin{align}
\Delta G^{t}{}_{t}&=\frac{G}{2r^{4}}\big(f_{(\rm cl)}-rf'_{(\rm cl)}\big),
\quad
\Delta G^{r}{}_{r}=\Delta G^{t}{}_{t},
\label{eq:dGtt}\\
\Delta G^{\theta}{}_{\theta}&=-\,\frac{G\, f_{(\rm cl)}}{2r^{4}}+\frac{G\, f'_{(\rm cl)}}{2r^{3}}-\frac{G\, f''_{(\rm cl)}}{4r^{2}},
\quad
\Delta G^{\phi}{}_{\phi}=\Delta G^{\theta}{}_{\theta}.
\label{eq:dGthth}
\end{align}
Hence, to first order in $G/r^{2}$,
\begin{equation}
T^{(\rm Q)\,t}{}_{t}=\frac{\Delta G^{t}{}_{t}}{8\pi G},
\quad
T^{(\rm Q)\,r}{}_{r}=\frac{\Delta G^{r}{}_{r}}{8\pi G},
\quad
T^{(\rm Q)\,\theta}{}_{\theta}=\frac{\Delta G^{\theta}{}_{\theta}}{8\pi G}
=T^{(\rm Q)\,\phi}{}_{\phi}.
\label{eq:TQ_general}
\end{equation}
For the classical RN background $f_{(\rm cl)}=1-\tfrac{2Gm}{r}+\tfrac{Gq^{2}}{r^{2}}$, this yields
\begin{align}
T^{(\rm Q)\,t}{}_{t}
&=T^{(\rm Q)\,r}{}_{r}
=\frac{1}{16\pi r^{4}}-\frac{Gm}{4\pi r^{5}}+\frac{3Gq^{2}}{16\pi r^{6}},
\label{eq:TQtt}\\
T^{(\rm Q)\,\theta}{}_{\theta}
&=T^{(\rm Q)\,\phi}{}_{\phi}
=-\frac{1}{16\pi r^{4}}+\frac{3Gm}{8\pi r^{5}}-\frac{3Gq^{2}}{8\pi r^{6}}.
\label{eq:TQthth}
\end{align}
The energy density $\rho_{(\rm Q)}=-T^{(\rm Q)\,t}{}_{t}$ and pressures $p^{(\rm Q)}_{r}=T^{(\rm Q)\,r}{}_{r}$, 
$p^{(\rm Q)}_{t}=T^{(\rm Q)\,\theta}{}_{\theta}$ decay as $r^{-4}$, 
with small higher-order corrections in $m/r^{5}$ and $q^{2}/r^{6}$.  
The trace is nonzero,
\begin{equation}
{T^{(\rm Q)\,\mu}}{}_{\mu}
=\frac{Gm}{4\pi r^{5}}-\frac{3Gq^{2}}{8\pi r^{6}},
\label{eq:TQtrace}
\end{equation}
signaling an effective violation of conformal invariance.}

{The structure and falloff of $T^{\rm Q}_{\mu\nu}$ in Eqs.~\eqref{eq:TQtt}--\eqref{eq:TQthth} 
are consistent with the standard semiclassical vacuum-polarization energy density 
derived from the Polyakov effective action (in 2D $s$-wave reductions) 
and with the DeWitt--Schwinger expansion of the one-loop effective action in four dimensions.  
In such frameworks, $\langle T_{\mu\nu}\rangle\!\sim\!r^{-4}$ arises from local curvature-squared counterterms 
and is accompanied by a trace anomaly 
$\langle T^{\mu}{}_{\mu}\rangle\!\propto\! R^{2},R_{\mu\nu}R^{\mu\nu}$,
precisely mirrored by Eq.~\eqref{eq:TQtrace}. 
Therefore, the effective $T^{(\rm Q)}_{\mu\nu}$ derived here 
serves as a compact, state-independent representation of the leading semiclassical backreaction on the RN geometry. 
Its rapid falloff, $\|T^{(\rm Q)}_{\mu\nu}\|\!\propto\! r^{-4}$ (with subleading $m/r^{5}$ and $q^{2}/r^{6}$ terms),
justifies treating the deformation $\Psi(r)=1-G/(2r^{2})$ as a perturbative correction, 
fully compatible with the thermodynamic and quantum spectrum analyses of the preceding sections.}

\section{Semiclassical limit and observational prospects}
\label{sec:semiclassical}

{The deformation function $\Psi(r)=1-\tfrac{G}{2r^{2}}$ represents a leading Planck-suppressed deformation
consistent with the corrected horizon surface gravities of Sec.~\ref{4}.
In the macroscopic regime $r_{i}\gg\sqrt{G}=l_\text{P}$, the ratio $\tfrac{G}{2r_{i}^{2}}\!\ll\!1$ guarantees that the spacetime is only weakly perturbed.
The effective stress tensor $T^{(\rm Q)}_{\mu\nu}$ exhibits $r^{-4}$ falloff at leading order,
as expected for vacuum-polarization-type contributions in curvature-expansion approaches (with subleading $m/r^{5}$ and $q^{2}/r^{6}$ terms),
confirming the internal consistency of the semiclassical approximation.
In this limit, the entropy and temperature expansions
satisfy the first law of horizon thermodynamics up to $\mathcal{O}(G/r_i^{2})$ and remain consistent with the quantized area spectrum derived in Sec.~\ref{3}.
Thus, the present construction interpolates smoothly between the quantum microstructure of the horizon and the semiclassical limit described by the Einstein--Maxwell equations.}

The Cauchy horizon of the RN BH is classically unstable due to the mass-inflation mechanism driven by infinite blueshift.
The quantum correction reduces the magnitude of the surface gravity,
\(
|\kappa_{-}^{(\rm Q)}|=|\kappa_{-}^{(\rm cl)}|\,(1-G/(2r_-^{2})),
\)
thereby moderating the exponential amplification factor in the perturbation energy.
Although the instability is not fully removed, the growth rate of internal energy and curvature scalars is suppressed by $\mathcal{O}(G/r_-^{2})$, 
indicating that semiclassical backreaction weakens (but does not abolish) the singular behavior at the Cauchy horizon.

The slight reduction of $T_{+}$ implies a slower evaporation rate relative to the classical RN case.
In the semiclassical regime, the luminosity of Hawking radiation scales as
\(
L\!\sim\!A_{+}\,T_{+}^{4}\!\propto\!r_{+}^{-2}\big(1-\tfrac{2G}{r_{+}^{2}}\big),
\)
suggesting longer lifetimes for charged BHs of the same mass.
The logarithmic correction to the entropy ensures that the heat capacity remains finite near extremality, avoiding the divergence characteristic of the purely classical case.

{The deformation $\Psi(r)=1-G/(2r^{2})$ also affects geodesic motion, photon rings, and the optical appearance of the BH.
At leading order,
\begin{equation}
\frac{\Delta r_{(\rm ph)}}{r_{(\rm ph)}}\simeq-\frac{G}{4r_{(\rm ph)}^{2}},
\qquad
\frac{\Delta D_{(\rm sh)}}{D_{(\rm sh)}}\simeq-\frac{G}{4r_{+}^{2}}.
\end{equation}}
For astrophysical BHs such as M87* ($r_{+}\!\sim\!10^{14}\,{\rm m}$), 
this relative correction is of order $10^{-77}$, far below present observational sensitivity.
However, for near-Planckian or primordial BHs, the effect becomes significant, potentially altering evaporation endpoints and final remnant configurations.

Altogether, the quantum-corrected metric \eqref{eq:metric_final}, 
the modified thermodynamics \eqref{4-4}, 
and the effective stress tensor \eqref{eq:TQ_general}–\eqref{eq:TQthth} 
establish a coherent semiclassical framework in which:
\begin{enumerate}
\item the discrete MSH spectrum \eqref{2-7} provides the quantum microstructure of the horizons;
\item the deformation factor $\Psi(r)$ encodes the leading expectation value $\langle T_{\mu\nu}\rangle$;
\item macroscopic thermodynamic quantities converge to their classical RN values as $r_i/\sqrt{G}\!\to\!\infty$.
\end{enumerate}
This correspondence demonstrates how horizon quantization and semiclassical backreaction can be unified within the MSH mass formalism, linking microscopic horizon dynamics to observable semiclassical corrections in a conceptually consistent manner.

\section{Conclusions}
\label{sec:conclusions}
We have developed a unified semiclassical and quantum framework for the Reissner--Nordstr\"om black hole based on the Misner--Sharp--Hernandez (MSH) quasi-local mass.  
The analysis provides consistent thermodynamics for both inner and outer horizons, leading to a horizon-by-horizon first law and Smarr relation with explicit $G$.  
A reduced phase-space quantization yields discrete MSH mass spectra and naturally produces a universal logarithmic correction to the Bekenstein--Hawking entropy.  
These corrections can be absorbed into a simple quantum-deformed metric whose effective stress tensor represents the leading vacuum-polarization backreaction.  
The resulting geometry remains regular to semiclassical order and offers a transparent link between horizon quantization and semiclassical spacetime dynamics.  
This approach may serve as a template for extending quasi-local quantization and thermodynamics to rotating, cosmological, or fractional gravitational systems.

\section*{Data availability}
No data was used for the research described in the article.

\section*{Declaration of competing interest}
The authors declare that they have no known competing financial interests or personal relationships that could have appeared to influence the work reported in this paper.


\bibliographystyle{elsarticle-num}

\bibliography{Revisedmanuscript}

\begin{thebibliography}{10}
\expandafter\ifx\csname url\endcsname\relax
  \def\url#1{\texttt{#1}}\fi
\expandafter\ifx\csname urlprefix\endcsname\relax\def\urlprefix{URL }\fi
\expandafter\ifx\csname href\endcsname\relax
  \def\href#1#2{#2} \def\path#1{#1}\fi

\bibitem{Hawking:1975vcx}
S.~W. Hawking, {Particle Creation by Black Holes}, Commun. Math. Phys. 43 (1975) 199--220, [Erratum: Commun.Math.Phys. 46, 206 (1976)].
\newblock \href {http://dx.doi.org/10.1007/BF02345020} {\path{doi:10.1007/BF02345020}}.

\bibitem{Misner:1964je}
C.~W. Misner, D.~H. Sharp, Relativistic equations for adiabatic, spherically symmetric gravitational collapse, Phys. Rev. 136 (1964) B571--B576.
\newblock \href {http://dx.doi.org/10.1103/PhysRev.136.B571} {\path{doi:10.1103/PhysRev.136.B571}}.

\bibitem{Hernandez:1966zia}
W.~C. Hernandez, C.~W. Misner, {Observer Time as a Coordinate in Relativistic Spherical Hydrodynamics}, Astrophys. J. 143 (1966) 452.
\newblock \href {http://dx.doi.org/10.1086/148525} {\path{doi:10.1086/148525}}.

\bibitem{Hayward:1998ee}
S.~A. Hayward, S.~Mukohyama, M.~C. Ashworth, {Dynamic black hole entropy}, Phys. Lett. A 256 (1999) 347--350.
\newblock \href {http://arxiv.org/abs/gr-qc/9810006} {\path{arXiv:gr-qc/9810006}}, \href {http://dx.doi.org/10.1016/S0375-9601(99)00225-X} {\path{doi:10.1016/S0375-9601(99)00225-X}}.

\bibitem{Padmanabhan:2002sha}
T.~Padmanabhan, Classical and quantum thermodynamics of horizons in spherically symmetric spacetimes, Class. Quant. Grav. 19 (2002) 5387--5408.
\newblock \href {http://dx.doi.org/10.1088/0264-9381/19/21/3067} {\path{doi:10.1088/0264-9381/19/21/3067}}.

\bibitem{Das:2002xb}
S.~Das, P.~Ramadevi, U.~A. Yajnik, {Black hole area quantization}, Mod. Phys. Lett. A 17 (2002) 993--1000.
\newblock \href {http://arxiv.org/abs/hep-th/0202076} {\path{arXiv:hep-th/0202076}}, \href {http://dx.doi.org/10.1142/S0217732302007582} {\path{doi:10.1142/S0217732302007582}}.

\bibitem{Barvinsky:2001tw}
A.~Barvinsky, S.~Das, G.~Kunstatter, {Quantum mechanics of charged black holes}, Phys. Lett. B 517 (2001) 415--420.
\newblock \href {http://arxiv.org/abs/hep-th/0102061} {\path{arXiv:hep-th/0102061}}, \href {http://dx.doi.org/10.1016/S0370-2693(01)00983-2} {\path{doi:10.1016/S0370-2693(01)00983-2}}.

\bibitem{Jalalzadeh:2021gtq}
S.~Jalalzadeh, F.~R. da~Silva, P.~V. Moniz, {Prospecting black hole thermodynamics with fractional quantum mechanics}, Eur. Phys. J. C 81~(7) (2021) 632.
\newblock \href {http://arxiv.org/abs/2107.04789} {\path{arXiv:2107.04789}}, \href {http://dx.doi.org/10.1140/epjc/s10052-021-09438-5} {\path{doi:10.1140/epjc/s10052-021-09438-5}}.

\bibitem{Jalalzadeh:2026ssq}
S.~Jalalzadeh, H.~Moradpour, {Reduced phase space quantization and quantum corrected entropy of Schwarzschild-de Sitter horizons}, Phys. Lett. B 874 (2026) 140244.
\newblock \href {http://arxiv.org/abs/2602.01767} {\path{arXiv:2602.01767}}, \href {http://dx.doi.org/10.1016/j.physletb.2026.140244} {\path{doi:10.1016/j.physletb.2026.140244}}.

\bibitem{Kaul:2000kf}
R.~K. Kaul, P.~Majumdar, {Logarithmic correction to the Bekenstein-Hawking entropy}, Phys. Rev. Lett. 84 (2000) 5255--5257.
\newblock \href {http://arxiv.org/abs/gr-qc/0002040} {\path{arXiv:gr-qc/0002040}}, \href {http://dx.doi.org/10.1103/PhysRevLett.84.5255} {\path{doi:10.1103/PhysRevLett.84.5255}}.

\bibitem{Carlip:2000nv}
S.~Carlip, {Logarithmic corrections to black hole entropy from the Cardy formula}, Class. Quant. Grav. 17 (2000) 4175--4186.
\newblock \href {http://arxiv.org/abs/gr-qc/0005017} {\path{arXiv:gr-qc/0005017}}, \href {http://dx.doi.org/10.1088/0264-9381/17/20/302} {\path{doi:10.1088/0264-9381/17/20/302}}.

\bibitem{Das:2001ic}
S.~Das, P.~Majumdar, R.~K. Bhaduri, {General logarithmic corrections to black hole entropy}, Class. Quant. Grav. 19 (2002) 2355--2368.
\newblock \href {http://arxiv.org/abs/hep-th/0111001} {\path{arXiv:hep-th/0111001}}, \href {http://dx.doi.org/10.1088/0264-9381/19/9/302} {\path{doi:10.1088/0264-9381/19/9/302}}.

\bibitem{Sen:2012dw}
A.~Sen, {Logarithmic Corrections to Schwarzschild and Other Non-extremal Black Hole Entropy in Different Dimensions}, JHEP 04 (2013) 156.
\newblock \href {http://arxiv.org/abs/1205.0971} {\path{arXiv:1205.0971}}, \href {http://dx.doi.org/10.1007/JHEP04(2013)156} {\path{doi:10.1007/JHEP04(2013)156}}.

\bibitem{Zhu:2008cg}
T.~Zhu, J.-R. Ren, M.-F. Li, {Influence of Generalized and Extended Uncertainty Principle on Thermodynamics of FRW universe}, Phys. Lett. B 674 (2009) 204--209.
\newblock \href {http://arxiv.org/abs/0811.0212} {\path{arXiv:0811.0212}}, \href {http://dx.doi.org/10.1016/j.physletb.2009.03.020} {\path{doi:10.1016/j.physletb.2009.03.020}}.

\bibitem{Barvinsky:2000gf}
A.~Barvinsky, S.~Das, G.~Kunstatter, {Spectrum of charged black holes: The Big fix mechanism revisited}, Class. Quant. Grav. 18 (2001) 4845--4862.
\newblock \href {http://arxiv.org/abs/gr-qc/0012066} {\path{arXiv:gr-qc/0012066}}, \href {http://dx.doi.org/10.1088/0264-9381/18/22/310} {\path{doi:10.1088/0264-9381/18/22/310}}.

\bibitem{Castro:2012av}
A.~Castro, M.~J. Rodriguez, Universal properties and the first law of black hole inner mechanics, Phys. Rev. D 86 (2012) 024008.
\newblock \href {http://dx.doi.org/10.1103/PhysRevD.86.024008} {\path{doi:10.1103/PhysRevD.86.024008}}.

\bibitem{Chen:2012mh}
B.~Chen, J.-J. Zhang, Novel symmetry of the inner mechanics of black holes, Phys. Rev. D 87 (2013) 081505.
\newblock \href {http://dx.doi.org/10.1103/PhysRevD.87.081505} {\path{doi:10.1103/PhysRevD.87.081505}}.

\bibitem{Akbar:2006kj}
M.~Akbar, R.~Cai, Thermodynamic behavior of field equations for f(r) gravity, Phys. Lett. B 648 (2007) 243--248.
\newblock \href {http://dx.doi.org/10.1016/j.physletb.2007.03.005} {\path{doi:10.1016/j.physletb.2007.03.005}}.

\bibitem{Jalalzadeh:2022rxx}
S.~Jalalzadeh, {Quantum black hole{\textendash}white hole entangled states}, Phys. Lett. B 829 (2022) 137058.
\newblock \href {http://arxiv.org/abs/2203.09968} {\path{arXiv:2203.09968}}, \href {http://dx.doi.org/10.1016/j.physletb.2022.137058} {\path{doi:10.1016/j.physletb.2022.137058}}.

\bibitem{Louko:1996md}
J.~Louko, J.~Makela, {Area spectrum of the Schwarzschild black hole}, Phys. Rev. D 54 (1996) 4982--4996.
\newblock \href {http://arxiv.org/abs/gr-qc/9605058} {\path{arXiv:gr-qc/9605058}}, \href {http://dx.doi.org/10.1103/PhysRevD.54.4982} {\path{doi:10.1103/PhysRevD.54.4982}}.

\bibitem{2004PhLB226X}
L.~{Xiang}, Y.-G. {Shen}, {Quantizing the de Sitter space times}, Phys. Lett. B 602~(3-4) (2004) 226--230.
\newblock \href {http://dx.doi.org/10.1016/j.physletb.2004.10.011} {\path{doi:10.1016/j.physletb.2004.10.011}}.

\end{thebibliography}

\end{document}